\documentclass[preprint,superscriptaddress,showkeys,showpacs,tightenlines,nofootinbib]{revtex4} 
\usepackage{graphicx}
\usepackage{epstopdf}
\usepackage{bm}
\begin{document}      
\preprint{PNU-NTG-06/2007}
\preprint{PNU-NuRI-06/2007}
\preprint{YITP-07-46}
\preprint{RCNP-Th07007}
\preprint{INHA-NTG-04/2008}
\title{$\eta$ photoproduction and $N^*$ resonances}
%---------------------------------------------------
\author{Ki-Seok Choi}
\email[E-mail: ]{kschoi@pusan.ac.kr}
\affiliation{Department of
Physics and Nuclear physics \& Radiation Technology Institute (NuRI),
Pusan National University, Busan 609-735, Korea} 
%---------------------------------------------------
\author{Seung-il Nam}
\email[E-mail: ]{sinam@yukawa.kyoto-u.ac.jp}
\affiliation{Yukawa Institute for Theoretical Physics (YITP), Kyoto
University, Kyoto 606-8502, Japan} 
%---------------------------------------------------
\author{Atsushi Hosaka}
\email[E-mail: ]{hosaka@rcnp.osaka-u.ac.jp}
\affiliation{Research Center for Nuclear Physics (RCNP), Ibaraki, Osaka
567-0047, Japan}
%----------------------------------------------------
\author{Hyun-Chul Kim}
\email[E-mail: ]{hchkim@inha.ac.kr}
\affiliation{Department of Physics, Inha University, Incheon 402-751, Korea} 
%-----------------------------------------------------
\date{March, 2008}
%-----------------------------------------------------
\begin{abstract}
We investigate the $\eta$ photoproduction using the effective
Lagrangian approach at the tree level.  We focus on the new nucleon
resonance $N^*(1675)$, which was reported by the GRAAL, CB-ELSA and
Tohoku LNS, testing its possible spin and parity states theoretically
($J^P=1/2^{\pm},3/2^{\pm}$).  In addition, we include six nucleon
resonances, $D_{13}(1520)$, $S_{11}(1535)$, $S_{11}(1650)$,
$D_{15}(1675)$, $P_{11}(1710)$, $P_{13}(1720)$ as well as the possible
background contributions.  We calculate various cross sections
including beam asymmetries for the neutron and proton targets.  We find
noticeable isospin asymmetry in transition amplitudes for photon and
neutron targets. This observation may indicate that the new resonance
can be identified as a non-strangeness member of the baryon
antidecuplet. 
\end{abstract} 
%-------------------------------------------------------
\pacs{13.75.Cs, 14.20.-c}
\keywords{$\eta$ photoproduction, GRAAL experiment, Non-strangeness
pentaquark} 
\maketitle
%-------------------------------------------------------
\section{Introduction}
%-------------------------------------------------------
After the first experimental report on the exotic baryon assigned as
$\Theta^+$ from the LESP collaboration at
SPring-8~\cite{Nakano:2003qx}, there have been a large number of
related experimental and theoretical works to date. Among them, the
GRAAL collaboration reported a new nucleon resonance $N^*(1675)$ from
$\eta$
photoproduction~\cite{Kuznetsov:2004gy,Kuznetsov:2006kt}\footnote{Recent
  analysis has shown that the mass of the new resonance is about
  $1.68$ GeV~\cite{Kuznetsov:2007gr}.}. Their data show a narrow peak
of which decay width $\Gamma_{N^*\to\eta N}$ was estimated to be about
$40$ MeV.  After the Fermi-motion correction being taken into account,
the width may become even narrower $\sim10$ MeV~\cite{Fix:2007st}.
This narrow width is a typical feature for the pentaquark exotic 
baryons~\cite{Diakonov:2003jj,Arndt:2003ga,Polyakov:2003dx}. Moreover, 
the production process of the $N^*(1675)$ largely depends on its
isospin state of the target nucleons: A larger $N^*(1675)$ peak is
shown for the neutron target, while it is suppressed for the proton
one.  Considering the fact that isospin-symmetry breaking is
negligible at the strongly interacting vertex, this large asymmetry
comes mainly from the photon coupling.  Interestingly,
$N^*(1675)$ being assumed as a member of the baryon antidecuplet
($\overline{10}$), this large isospin asymmetry was well explained in the
chiral quark-soliton model
($\chi$QSM)~\cite{Kim:2005gz,Azimov:2005jj}.  In fact, it originates
from the $U$-spin conservation in the photon
coupling~\cite{Hosaka}.  This assymetry was emphasized once again 
experimentally in Ref.~\cite{Kuznetsov:2007dy}. Recently, the Tohoku
LNS~\cite{Tohoku} and CB-ELSA~\cite{CBELSA} reported $\eta$
photoproduction from the deuteron target, providing the same
conclusion on that behavior. Concerning the spin and parity, their
assignments are not yet determined unambiguously. Although the
$\eta$-MAID has assumed $J^P=1/2^+$ as suggested by the
$\chi$QSM~\cite{Fix:2007st}, in our previous work~\cite{Choi:2005ki},
we have shown that $J^P=1/2^-$ was equally possible in comparison with
the experimental data.    

In the present work, following Ref.~\cite{Choi:2005ki}, we would
like to present a recent study on the $\eta$ photoproduction employing
the effective Lagrangian approach in the Born approximation.  We
include six nucleon resonances
$D_{13}(1520)$, $S_{11}(1535)$, $S_{11}(1650)$, $D_{15}(1675)$,
$P_{11}(1710)$, $P_{13}(1720)$
in a fully relativistic manner. We ignore the contributions from
$N^*(1680,5/2)$ and $N^*(1700,3/2)$ considered in Ref.~\cite{Fix:2007st} since
their branching ratios to the $\eta N$ channel are negligible.  Nucleon
pole terms and vector-meson exchanges are also taken into account as
backgrounds. In order to test the spin and  parity of the new
resonance, we investigate the different four cases,
$J^P=1/2^{\pm},3/2^{\pm}$. We utilize the phenomenological form
factors in terms of a gauge-invariant manner as done in 
Ref.~\cite{Choi:2005ki,Nam:2005uq,Nam:2005jz}. As a result, we observe
that $\mu_{\gamma{n}n^*(1675)}=0.1\sim0.2$ and
$\mu_{\gamma{p}p^*(1675)}\simeq0$ for $J^P=1/2^{\pm}$ whereas
$\mu_{\gamma nn^*(1675)}=0.01\sim0.02$ and $\mu_{\gamma
  pp^*(1675)}\simeq0$ for $J^P=3/2^{\pm}$ to reproduce the GRAAL data
qualitatively well. Here, $\mu_{\gamma{N}N^*(1675)}$ denotes the
strengths of either magnetic or electric coupling between $N$ and
$N^*(1675)$.  The present study seems to prefer the nucleon resonance
of $J^P=1/2^{\pm}$, but the possibility of higher spin can not be
completely excluded. 

The present work is organized as follows: In Section II, we provide
our formalisms for $\eta$ photoproduction.  Differential cross
sections and beam asymmetry being compared to the GRAAL data are given
in Section III and Section IV, respectively, with discussions.  Final
section is devoted for summary and conclusion. 
  
%-------------------------------------------------------
\section{Formalism}
\begin{figure}[t]
\includegraphics[width=12cm]{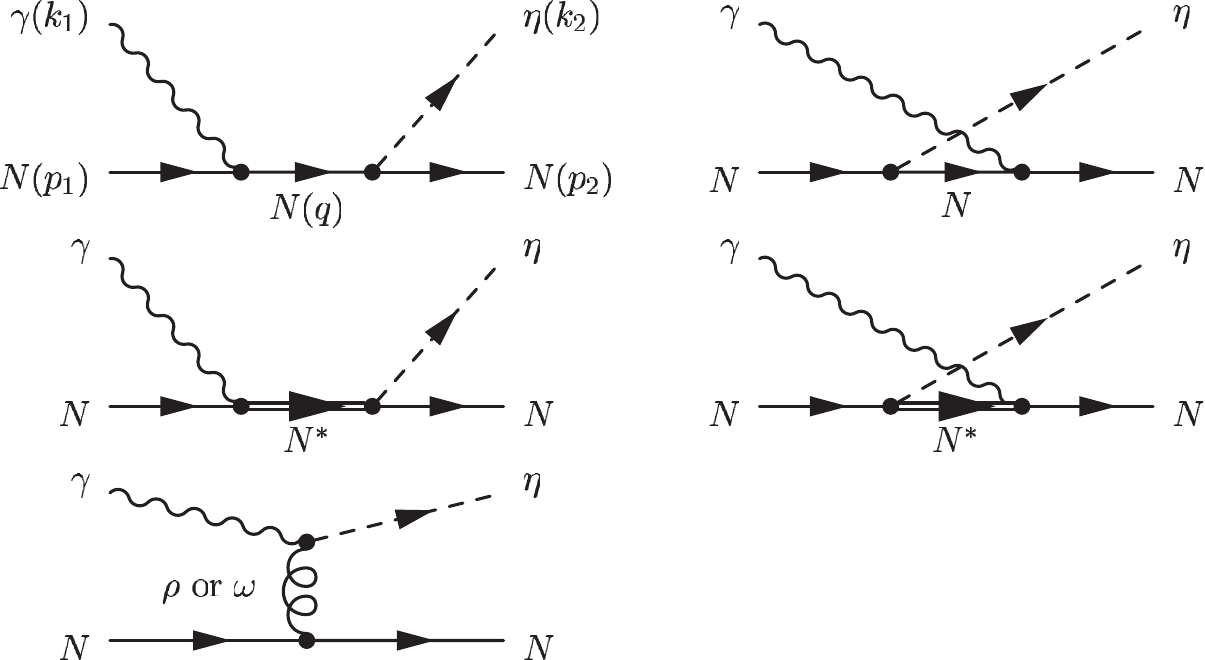}
\caption{Relevant tree-level diagrams for $\eta$ photoproduction.}     
\label{fig0}
\end{figure}

%-------------------------------------------------------
In this section, we present our model for $\eta$ photoproduction.  In 
Figure~\ref{fig0} we show the relevant tree-level diagrams for it
schematically.  The diagrams represent nucleon-pole (top) and resonance
(middle) contributions in $s$- (left) and $u$-channels (right). In
addition we also consider vector-meson exchanges ($\rho$ and $\omega$)
in $t$-channel (bottom).  
The nucleon-pole and  vector-meson exchange terms are assigned
as the background contributions. The effective
Lagrangians for the Yukawa vertices for the background contributions
can be written as follows:  
\begin{eqnarray}
 \mathcal{L}_{\gamma NN}
&=&-e\bar{N}\rlap{/}{A}N - i
\frac{e\kappa_N}{2M_N}\bar{N}\sigma_{\mu\nu}q^{\nu}A^{\mu}N+{\rm h.c.},
\nonumber\\
\mathcal{L}_{\eta NN}&=&-ig_{\eta NN}\bar{N}\gamma_5 \eta N+{\rm h.c.},
\nonumber \\
\mathcal{L}_{VNN}&=& -g^v_{VNN} \bar{N}\rlap{/}{V}N-
 i\frac{g^t_{VNN}}{2M_N}\bar{N}\sigma_{\mu\nu}q^{\nu}V^{\mu}N+{\rm h.c.}.,
\nonumber\\
\mathcal{L}_{ \gamma\eta V}&=&\frac{eg_{ \gamma\eta V}}{4M_{\eta}}
\epsilon_{\mu\nu\sigma\rho}F^{\mu\nu}V^{\sigma\rho}\eta+{\rm h.c.}, 
\label{backgrounds}
\end{eqnarray}
where $\gamma$, $N$, $\eta$ and $V$ stand  for the fields of photon, 
nucleon, $\eta$ meson and vector mesons ($\rho$ and $\omega$), 
respectively.  The $e$ and $\kappa_N$ denote the electric charge and
anomalous magnetic moment of the nucleon, respectively. $M_h$ denotes
the mass of the hadron $h$. The strength of the relevant couplings are
employed from the Nijmegen potential model~\cite{Stoks:1999bz} as
shown in Table~\ref{table0}.  
\begin{table}[b]
\begin{tabular}{c|c|c|c|c|c|c}
$g_{\eta NN}$&$g_{\rho NN}^v$&$g_{\rho NN}^t$&
$g_{\omega NN}^v$&$g_{\omega NN}^t$&$g_{\rho \eta \gamma}$&$g_{\omega 
  \eta \gamma}$\\ 
\hline
0.47&2.97&12.52&10.36&4.20&0.89&0.192\\
\end{tabular}
\caption{Strong and photon couplings for the background contributions.}
\label{table0}
\end{table}

In the following we present the effective Lagrangians for the resonant 
contributions of spin 1/2, 3/2 and 5/2:
\begin{eqnarray}
{\cal L}_{\gamma N N^*}^{1/2}&=& \frac{e\mu_{\gamma N N^*}
 }{2(M_N+M_{N^*})}\bar{N}^*\Gamma^a_5\sigma_{\mu\nu}F^{\mu\nu}N,
\nonumber\\ 
{\cal L}_{\gamma N
    N^*}^{3/2}&=&\frac{ie\mu_{\gamma{N}N^*}}{M_{N^*}}\bar{N}^{*\mu}
\Theta_{\mu\nu}(C,D)\Gamma^b_5\gamma_{\lambda}NF^{\lambda \nu},
\nonumber\\
{\cal L}_{\gamma N
    N^*}^{5/2}&=&\frac{e\mu_{\gamma{N}N^*}}{M^2_{N^*}}\bar{N}^{*\mu\alpha}
\Theta_{\mu\nu}(E,F)\gamma_{\lambda}\Gamma^a_5(\partial_{\alpha}
F^{\lambda\nu})N,
\nonumber\\
 {\cal L}_{\eta N N^*}^{1/2}&=&-ig_{\eta N N^*} \bar{N }
\Gamma \gamma_5 \eta N^*,
\nonumber\\
{\cal L}_{\eta N N^*}^{3/2}&=&\frac{g_{\eta N
    N*}}{M_{\eta}}\bar{N}^{*\mu}\Theta_{\mu\nu}(A,B)\Gamma^a_5 N,
    \partial^{\nu}\eta \nonumber \\
{\cal L}_{\eta N N^*}^{5/2}&=&\frac{g_{\eta N
    N^*}}{m^2_{\eta}}\bar{N}^{*\mu\nu}\Theta_{\mu\delta}(A,B)
\Theta_{\nu\lambda}(C,D)\Gamma^b_5N
    \partial^{\delta}\partial^{\lambda}\eta, 
\label{resonance}
\end{eqnarray}
where the spinors for spin-3/2 and spin-5/2 fermions are defined by the
Rarita-Schwinger formalism~\cite{Rarita:1941mf,Titov:2002iv} as given
in Appendix.  Moreover, as for the Lagrangians with the higher spins
(3/2 and 5/2), it is necessary to take into account the off-shell
parameter due to the point-transformation
invariance~\cite{Read:1973ye}:  
\begin{equation}
\Theta_{\mu\nu}(A,B)=g_{\mu\nu}+\left[\frac{1}{2}\left(1+4B\right)A+B\right] 
\gamma_{\mu}\gamma_{\nu}.
\label{offshell}
\end{equation}
Considering the gauge-invariance~\cite{Nath:1971wp}, we take $A=-1$,
we can rewrite Eq.~(\ref{offshell}) as follows:  
\begin{equation}
\Theta_{\mu\nu}(X)=g_{\mu\nu}+X\gamma_{\mu}\gamma_{\nu},
\,\,\,\,X=-\left(B+\frac{1}{2}\right).
\end{equation}
One can refer to the related topics for determining the off-shell
parmaters thoretically in Refs.~\cite{Kimel:1973qx,Nath:1971wp}.  In 
the present work, we set the off-shell parameters to be zero for
simplicity.  This choice, however, not bad, since resonance
contributions become important only near their on-mass shells.  The  
parity of the resonances are controlled by 
the matrices $\Gamma^a_5$ and $\Gamma^b_5$:
 \begin{eqnarray}
{\rm Positive\,\,parity}\, &:&\Gamma^a_5={\bf 1}_{4\times4},
\,\,\,\,\Gamma^b_5=\gamma_5,\nonumber\\ 
{\rm Negative\,\,parity}\, &:&\Gamma^a_5=\gamma_5,
\,\,\,\,\Gamma^b_5={\bf 1}_{4\times4}. 
\end{eqnarray}

We determine the strengths of the strong couplings for the
resonance of spin $n/2$ ($n=1,3,5$) using the decay width
$\Gamma_{N^* \to \eta N}$ as input:  
\begin{equation}
g^{n/2}_{\eta NN^*}=\left[\frac{2(n+1)\,\pi M_{N^*}M^{n-1}_{\eta}}{{\cal C}^2_n\,
|{\bm P}_f|^n\left(\sqrt{M^2_N+|{\bm P}_{\eta N}|^2}
-\Pi\sin\left(\frac{n\pi}{2}\right)M_N\right)}\,
\Gamma_{N^* \to \eta N}\right]^{\frac{1}{2}},
\label{strong}
\end{equation}
${\cal C}_n$ stands for the Clebsch-Gordan coefficient for the spin
transition $N\to N^*$ where ${\cal
  C}_{1,3,5}=1,\sqrt{2/3},\sqrt{2/5}$.  The three momentum of the
particles in the final state ($\eta$ and $N$), $|{\bm P}_{h_1h_2}|$ is
deinfed by   
\begin{equation}
|{\bm P}_{h_1h_2}|
=\left[\left(\frac{M_{N^*}^2-M_{h_1}^2
+M_{h_2}^2}{2M_{N^*}}\right)^2-M_{h_2}^2\right]^{\frac{1}{2}}.
\end{equation}

The photon coupling to the resonances ($\mu_{\gamma NN^*}$ ) can be
computed by~\cite{Titov:2002iv}:     
\begin{eqnarray}
\label{eeeee}
 e\mu^{1/2}_{\gamma NN^*}&=&\pm \left[\frac{M_{N^*}M_N}
{|{\bm P}_{\gamma N}|}\right]^{\frac{1}{2}}|A_{1/2}^{N^*}|,
\nonumber\\
e\mu^{3/2}_{\gamma NN^*}&=&\pm S^{N^*}_{3/2}
\sqrt{\frac{6M_NM^3_{N^*}}{\left(3M^2_{N^*}+M_N^2\right)
|{\bm P}_{\gamma N}|}}
\left[(A^{N^*}_{1/2})^2+A_{3/2}^{N^*})^2\right]^{1/2} 
\nonumber\\
e\mu^{5/2}_{\gamma NN^*} &=& \mp
  S^{N^*}_{5/2}\sqrt{\frac{10M_NM^5_{N^*}}
{\left(2M_{N^*}+M^2_N\right)|{\bm P}_{\gamma N}|^3}}
\left[(A^{N^*}_{1/2})^2+(A_{3/2}^{N^*})^2\right]^{1/2}.
\end{eqnarray}
Here, we account for the experimental fact that there is no strong 
correlation observed between $A^{3/2}$ and $A^{1/2}$.  Thus, we
ignored the interference between them in Eq.~(\ref{eeeee}). The sign
of the coupling is determined by the dominant contribution using the
following relations:  
\begin{eqnarray}
S_{n/2}^{N^*}&=
&{\rm sign}(A^{N^*}_{1/2})\,\theta(h^{N^*}_{n/2})
+{\rm sign}(A^{N^*}_{3/2})\,\theta(-h^{N^*}_{n/2}),
\nonumber\\
h^{N^*}_{3/2}&=&|A^{N^*}_{1/2}|-\frac{1}{\sqrt{2}}\frac{M_N}{M_{N^*}}
|A^{N^*}_{3/2}|,
\,\,\,\,
h^{N^*}_{5/2}=|A^{N^*}_{1/2}|-\frac{1}{\sqrt{3}}\frac{M_N}{M_{N^*}}
|A^{N^*}_{3/2}|,
\end{eqnarray}
where $\theta$ is the Heaviside step function.  In Table~\ref{table1}
we list all phenomenological input for the strong and electromagnetic
couplings for the relevant resonances in $1.7$ GeV
$\lesssim{E}_{\rm{cm}}\lesssim1.9$ GeV.  Note that all the strong
coupling strengths are determined in order to reproduce the total and
differential cross section data for the proton
target~\cite{Crede:2003ax} without the new resonance, since it was
already shown in our previous work~\cite{Choi:2005ki} that the
contribution from $p^*(1675)$ was negligible.  We also fixed the photon
couplings from the helicity amplitudes, listed in
Table~\ref{table1}~\cite{Yao:2006px}.  However, as for $F_{15}(1680)$,
we take relatively small values for the proton but large ones for the
neutron, respectively, in order to obtain our numerical results
compatible to the experimental data of
Ref.~\cite{Kuznetsov:2004gy,Kuznetsov:2006kt}.  Note that we 
have excluded $D_{13}(1700)$ considering its negligible branching
ratio decaying to $\eta{N}$. We note that the branching ratio 
for $D_{15}(1675)$ is much smaller than that used in the usual
$\eta$-MAID analyses $\sim17\%$~\cite{Fix:2007st}.  We use
$\Gamma_{N^*(1675)}\simeq40$ MeV and
$\Gamma_{N^*(1675)}/\Gamma_{N^*(1675)\to\eta N}\simeq0.25$ for the new 
resonance as done in the previous work~\cite{Choi:2005ki}.  This
choice is also equivalent to take the Fermi-motion correction into
account, resulting in the smearing of the decay width~\cite{Azimov}.  

\begin{table}[t]
\begin{tabular}{c||c|c|c|c|c|c}
$N^*$&$\Gamma_{N^*}$&$\Gamma_{N^*\to\eta N}/\Gamma_{N^*}$
&$A^{n^*}_{1/2}$& $A^{p^*}_{1/2}$&$A^{n^*}_{3/2}$&$A^{p^*}_{3/2}$\\
\hline
\hline
$D_{13}(1520)$&115&$2.3\times 10^{-3}$&$-0.059$&$-0.024$&$-0.139$&0.166\\
\hline
$S_{11}(1535)$&130 &55&$-0.043$&0.072 &$\cdots$&$\cdots$\\
\hline
$S_{11}(1650)$&150&3&$-0.041$&0.037&$\cdots$&$\cdots$\\
\hline
$D_{15}(1675)$&165&0.05&$-0.043$&0.011&$-0.058$&0.006\\
\hline
$N^*(1675)$&40&25&$\cdots$&$\cdots$&$\cdots$&$\cdots$\\
\hline
$F_{15}(1680)$&130&1&$0.080$&-0.009&$-0.040$&0.015\\
\hline
$D_{13}(1700)$&100 &0.025&$0.00$&-0.018&$-0.003$&$-0.002$\\
\hline
$P_{11}(1710)$&100 &6.2&$-0.002$&0.009&$\cdots$&$\cdots$\\
\hline
$P_{13}(1720)$&200&4 &0.001&0.010&$0.029$&$-0.011$ \\
\end{tabular}
\caption{Relevant inputs for numerical calculations: full decay width
  [MeV], branching ratio [$\%$] and helicity amplitude
  [GeV$^{-1/2}$]. The new nucleon resonance is denoted by
  $N^*(1675)$. }  
\label{table1}
\end{table}

The invariant amplitudes at the tree level can be evaluated
straightforwardly using the effective Lagrangians given in
Eqs.~(\ref{backgrounds}) and (\ref{resonance}):  
\begin{eqnarray}
\label{eq:amplitude}
i\mathcal{M}_s&=&\frac{eg_{\eta NN}}{s-M^2_N}
\bar{u}(p_2)\gamma_5\left[F^{N}_s
\rlap{/}{k}_1+F_c(\rlap{/}{p}_1+M_N)
+\frac{\kappa_NF^{N}_s}{2M_N}
\left(\rlap{/}{k}_1+\rlap{/}{p}_1+M_N\right)
\rlap{/}{k}_1\right]\rlap{/}{\epsilon}u(p_1),
\nonumber\\
i\mathcal{M}_u&=&\frac{eg_{\eta NN}}{u-M^2_N}
\bar{u}(p_2)\rlap{/}{\epsilon}\left[F_c(\rlap{/}{p}_2+M_N)-F^N_s
\rlap{/}{k}_1-\frac{\kappa_NF^{N}_u}{2M_N}
\rlap{/}{k}_1(\rlap{/}{p}_2-\rlap{/}{k}_1+M_N)
\right]\gamma_5u(p_1),
\nonumber\\
i\mathcal{M}_t&=&\frac{-ieg_{\gamma\eta{V}}F^V_t
\epsilon_{\mu\nu\sigma\rho}}{M_{\eta}\left(t-M^2_V\right)}
\bar{u}(p_2)k^{\mu}_1
\epsilon^{\nu}(k_1-k_2)^{\sigma}\left[g^v_{VNN}\gamma^{\rho}
+\frac{g^t_{VNN}}{4M_N}\left[\rlap{/}{q}\gamma^{\rho}-
\gamma^{\rho}(\rlap{/}{k}_1-\rlap{/}{p}_1)\right]\right]u(p_1),
\nonumber\\
i\mathcal{M}_{s^*}^{1/2}&=&\frac{e\mu_{\gamma N N^*} 
g_{\eta NN^*}F^{N^*}_s}
{(M_N+M_{N^*})\left[s-M^2_{N^*}-iM_{N^*}\Gamma_{N^*}\right]}
\bar{u}(p_2)\gamma_5 \Gamma^a_5
(\rlap{/}{k}_1+\rlap{/}{p}_1+M_{N^*}) \Gamma^a_5\rlap{/}{\epsilon}\rlap{/}{k}_1u(p_1),
\nonumber\\
i{\cal M}_{s^*}^{3/2}&=&-\frac{ieg_{\eta N N^*}\mu_{\gamma{N}N^*}F^{N^*}_s}
{M_{N^*}  M_{\eta}}\bar{u}(p_2)\Gamma^a_5 D^{3/2}_{\mu\nu}
k_2^{\mu}\Gamma^b_5 \gamma_{\sigma}(k_1^{\sigma}
\epsilon^{\nu}-k_1^{\nu}\epsilon^{\sigma})u(p_1)
 \nonumber \\
i{\cal
  M}_{s^*}^{5/2}&=&-\frac{eg_{\eta{N}N^*}\mu_{\gamma{N}N^*}F^{N^*}_s}{
M^2_{\eta}M^2_{N^*}}\bar{u}(p_2)\Gamma^b_5 
    k^{2\mu}k^{2\nu}D^{5/2}_{\mu\nu\rho\sigma}\gamma^{\lambda}
    \Gamma^a_5k^{1\rho} 
\left(k^{1\sigma} \epsilon_{\lambda}-k_{1\lambda}
    \epsilon^{\sigma}\right)u(p_1) 
\nonumber\\
  i{\cal M}_{u^*}^{3/2}&=&-\frac{ieg_{\eta N N^*}\mu_{\gamma{N}N^*}F^{N^*}_u}{M_{N^*}
  M_{\eta}}\bar{u}(p_2)\Gamma^b_5 \gamma_{\sigma}D^{3/2}_{\mu
  \nu}\left(k_1^{\sigma}\epsilon^{\mu}-k_1^{\mu}\epsilon^{\sigma}\right)
\Gamma^a_5 k_2^{\nu}u(p_1),
\nonumber\\
i\mathcal{M}_{u^*}^{1/2}&=&\frac{e\mu_{\gamma N N^*} 
g_{\eta NN^*}F^{N^*}_u}{(M_N+M_{N^*})\left[u-M^2_{N^*}-iM_{N^*}
\Gamma_{N^*}\right]} 
\bar{u}(p_2)\Gamma^a_5
\rlap{/}{\epsilon}\rlap{/}{k}_1(\rlap{/}{k}_2-\rlap{/}{p}_1+M_{N^*})
\gamma_5\Gamma^a_5 u(p_1),
\nonumber\\
i{\cal M}_{u^*}^{5/2}&=&-\frac{eg_{\eta{N}N^*}\mu_{\gamma{N}N^*}F^{N^*}_u}
{m^2_{\eta}M^2_{N^*}}\bar{u}(p_2)\gamma_{\lambda}
\Gamma^a_5 k_{1\nu} (k^{\mu}_1 \epsilon^{\lambda}-k^{\lambda}_1 
\epsilon^{\mu})D^{\mu\nu\sigma\rho} \Gamma^b_5 k_{2\sigma} 
k_{2\rho} u(p_1),
\end{eqnarray}
where the Mandelstam variables are defined as $s=(k_1+p_1)^2$,
$s=(k_2-p_1)^2$ and $t=(k_1-k_2)^2$. $D^{3/2}_{\mu\nu}$ and
$D^{5/2}_{\mu\nu\sigma\rho}$ denote the spin-3/2 and-5/2 propagators,
respectively, in terms of the projections of the Rarita-Schwinger
spinors (see Appendix):   
\begin{eqnarray}
  D^{3/2}_{\mu\nu}&\simeq&\frac{\rlap{/}{q}+M_{N^*}}{q^2-M^2_{N^*}-iM_{N^*}
\Gamma_{N^*}}\,g_{\mu\nu},
\nonumber\\
D^{5/2}_{\mu\nu\sigma\rho}&\simeq&\frac{\rlap{/}{q}+M_{N^*}}{q^2-M^2_{N^*}-iM_{N^*}
\Gamma_{N^*}}
\left[\frac{1}{2}(T_{\mu\sigma}T_{\nu\rho}+T_{\mu\rho}
T_{\nu\sigma})\right],
\nonumber\\
T^{\mu\nu}&=&-g^{\mu\nu}+\frac{q^{\mu}q^{\nu}}{M_{N^*}^2}.
\label{propagator}
\end{eqnarray}
where $q$ designates the momentum of the resonance.  As indicated in
Ref.~\cite{Nam:2005uq,Nam:2005jz}, we use the simplified form for the
spin-3/2 propagator as given in Eq.~(\ref{propagator}), 
since this form dominates the low energy region.  Similarly, as for
the spin-5/2 propagator, we consider the dominant contribution only. 

Considering the extended structure of the hadrons ($h$), we introduce
a phenomenological form factor for the resonance and background
contributions in a gauge-invariant
manner~\cite{Ohta:1989ji,Haberzettl:1998eq,Davidson:2001rk,Nam:2004xt}:   
\begin{eqnarray}
F^h_x=\left[\frac{\Lambda^4}{\Lambda^4+\left(x-M^2_h\right)^2}\right]^m,
\label{formfactor}
\end{eqnarray} 
where $x$ denotes the kinematical channel as well as the Mandelstem
variables ($s,u,t$). The power $m$ is chosen to be unity for the
spin-1/2 and -3/2 cases, and two for the spin-5/2 ones. This choice
enables us to reduce unphysical behavior (monotonically increasing
behavior beyond $E_{\rm{cm}}\sim{M}_{N^*}$) from the spin-5/2
propagator~\footnote{We note that there have been several methods to
  remedy this unitarity breaking problem including the Blatt-Weisskopf
  penetration factor and introducing phenomenological off-shell
  parameters~\cite{Titov:2002iv,Post:2000qi,Manley:1992yb,Manley:1984jz}.}.
To satisfy gauge invariance and normalization condition of the form
factor, we parameterize the form factor $F_c$ as
follows~\cite{Ohta:1989ji,Haberzettl:1998eq,Davidson:2001rk,Nam:2004xt}:  
\begin{eqnarray}
F_c=F^h_s+F^h_u-F^h_sF^h_u.
\label{commonformfactor}
\end{eqnarray}
We note that the amplitudes for the vector-meson exchange and
resonant contributions are gauge invariant. The cuf-off masses 
are determined such that the data are well produced: 
\begin{equation}
\Lambda_{N^*}=0.9\,{\rm GeV},\,\,
\Lambda_{N}=0.8\,{\rm GeV},\,\,
\Lambda_{\rho}=0.8\,{\rm GeV},\,\,
\Lambda_{\omega}=1.2\,{\rm GeV},\,\,
\end{equation}
%-------------------------------------------------------
\section{Numerical results}
%-------------------------------------------------------
In this Section, we provide numerical results for total and
differential cross sections and beam asymmetries. First, we show those
for the total cross sections as functions of the photon energy in the
laboratory frame in Fig.~2 for the proton target (first and second
rows) and neutron target (third and fourth rows).  The results are
shown separately for each spin-parity combination and the sign of 
$\mu_{\gamma NN^*}$ ($J^P_\mathrm{sign}=1/2^\pm_\pm$ and
$3/2^\pm_\pm$) . When we turn off the effects from the new resonance
($\mu_{\gamma pp^*}\approx0$), the proton data are well
reproduced. This tendency is  well compatible with estimations from 
the $\chi$QSM~\cite{Kim:2005gz} and consistent with our previous
results~\cite{Choi:2005ki}.  On the contrary, as for the neutron
target, we have varieties in the shapes of the curves due to the
interferences between the new resonance and 
other contributions, depending on its spin, parity and
$\mathrm{sign}(\mu_{\gamma nn^*})$. We observe that there is
constructive interference for the cases of
$J^{P}_\mathrm{sign}=1/2^+_+$, $1/2^-_-$ and $3/2^+_+$, showing the
peak due to the new resonance $N^*(1675)$. Note that the interference
is mainly due to the new resonance and the spin-5/2 resonances,
$F_{15}(1680)$ and $D_{15}(1675)$.  The absolute values of the photon
coupling are found to be $|\mu_{\gamma nn^*}|\ge0.1$ for the spin-1/2
cases and $|\mu_{\gamma nn^*}|\ge0.01$ for the spin-3/2 ones,
respectively, to provide clear peaks, indicating the evidence for the
new resonance.   

In Figs.~3 and 4, we draw the results for the differential cross
sections as functions of $\cos\theta$, in which $\theta$ stands for
the angle between the incident photon and outgoing kaon in the center
of mass (CM) frame, at a fixed photon energy $E_\gamma=800$ and $1145$
MeV, where experimental data are available.  At $E_\gamma\approx800$
MeV, which is almost the same with the threshold value,  we have
rather flat curves dominated by $s$-wave contributions as shown in
Fig.~3.  Especially, we can reproduce well the experimental data, taken
from Refs.~\cite{Crede:2003ax,Bartalini:2007fg}, for the proton target
case.  The neutron results are also dominated by the $s$-wave ones and
do not depend much on the spin and parity of the new resonance. As the
photon energy increases, higher wave contributions start to come into
play as shown in Fig.~4 for $E_\gamma\approx1145$ MeV.  Again, the
proton data are well reproduced.  However, we have rather different
shapes for the neutron showing a bump structure in the backward
scattering region $-1\le\cos\theta\le-0.5$.  We verified that these
bumps are due to the $F_{15}(1680)$ contribution. Note that the
dependence on $J^P_\mathrm{sign}$ is not so obvious again for all the
spin-parity cases.  

In Figs.~5 and 6, We show the differential cross sections with respect
to the center of mass (CM) energy at fixed angle $\theta=140$ and $65$
degrees, respectively. The experimental data for the neutron and
proton targets are taken from Ref.~\cite{Kuznetsov:2007gr}. Our
results for the proton target slightly deviate from the data at
$\theta\approx140^\circ$ as shown in Fig.~7, although the energy
dependence beyond $E_\mathrm{CM}\approx1750$ MeC, as well as the total
cross section at the entire energy region are reproduced well. In
contrast, the neutron data are well reproduced for the cases of
$J^{P}_\mathrm{sign}=1/2^+_+$, $1/2^-_-$ and $3/2^+_+$ as in the cases
for the total cross section results. It turns out that $|\mu_{\gamma
  nn^*}|\approx0.1\sim0.2$ for the spin-1/2 cases and $|\mu_{\gamma
  nn^*}|\approx0.01\sim0.02$ for the spin-3/2 ones are appropriate to
reproduce the observed peak. It is worth mentioning that, as for the
$J^P=1/2^+$ case, the estimated value of $|\mu_{\gamma nn^*}|$ is
consistent with the result of the
$\chi$QSM~\cite{Kim:2005gz}. Focusing in the forward scattering region
$\theta\approx65^\circ$ as shown in Fig.~6, our results for the proton
target are overestimated in comparison to the experimental
data. Interestingly, since the effects of $F_{15}(1680)$ and
$D_{15}(1675)$ become stronger for the neutron target at this angle,
the peak around $E_\mathrm{CM}\approx1.7$ MeV becomes as high as that
of $S_\mathrm{11}(1535)$. Thus, it is not easy to find a clear
evidence for the new resonance in the forward scattering region.     

Finally, we would like to discuss the beam asymmetry for the
present reaction process, which is defined as follows: 
%EQUATION>>>
\begin{equation}
\Sigma=\left[\frac{d\sigma}{d\Omega}_{\parallel}
-\frac{d\sigma}{d\Omega}_{\perp}\right]\times\left[\frac{d\sigma}
{d\Omega}_{\parallel}+\frac{d\sigma}{d\Omega}_{\perp}\right]^{-1}.
\label{eq:BA}
\end{equation}
%EQUATION<<<
Here the subscript $\parallel$ denotes that the polarization vector of 
the incident photon is parallel to the reaction plane, and vice versa
for $\perp$. From this definition, electric dominance of the photon
coupling ($E$) gives $\Sigma\sim+1$ whereas magnetic one ($M$) does
$\Sigma\sim-1$. We show our results in Figs.~7 and 8 as functions of
$\theta$ at a fixed photon energy $E_\gamma=870$ and $1051$ MeV for
the same $J^P_\mathrm{sign}$. In the vicinity of the threshold
($E_\gamma\approx870$ MeV), the proton data are well reproduced
showing a positive bump at around $\theta=90^\circ$ which indicates
that the electric photon coupling prevails. The bump structures are
slightly shifted to the forward angle for the neutron target case,
because of the effects from the spin-5/2 resonance
contributions. However, from the beam asymmetry results, we can not
see clear evidence for the new resonance. As the energy grows
($E_\gamma\approx1051$ MeV), the numerical results start to deviate
from the experimental data for the proton as shown in Fig.~8; the
theory is almost symmetric, while the experiment asymmetric and entitled
to the forward scattering region.  We consider that this deviation may 
result from unknown resonances, which are not taken into account 
here. Interestingly, although the results for the neutron are all
similar to each  $J^P_\mathrm{sign}$ case, the beam asymmetry becomes
negative, resulting from the strong interference between the new 
resonance and $F_{15}(1680)$.  

%-------------------------------------------------------
\section{Summary and Conclusion}
%-------------------------------------------------------
We have investigated the $\eta$-meson photoproduction in an effective
Lagrangian approach where the scattering amplitude was computed in the
Born approximation. Following our previous work, we studied the role
of the new nucleon resonance at round $E_{\rm{cm}}\sim1.675$ GeV
testing its possible spin and parity theoretically, considering
$1/2^{\pm}$ as well as $3/2^{\pm}$. In addition to this resonance, we
considered six other nucleon resonances, i.e. ($D_{13}(1520)$,
$S_{11}(1535)$, $S_{11}(1650)$, $D_{15}(1675)$, 
$P_{11}(1710)$, $P_{13}(1720)$), and nucleon-pole and vector-meson
exchange contributions as backgrounds. All coupling strengths were
determined by available experimental data and those obtained in the
Nijmegen potential.  Gauge invariance was satisfied explicitly via
appropriate form factor schemes.  

Total and differential cross sections were computed and compared with
the experimental data. In generl, the data for the proton were well
reproduced without the new resonance contribution, while for the
neutron data with that for $J^+_\mathrm{sign}=1/2^+_+$, $1/2^-_-$ and
$3/2^+_-$. Moreover, it turned out that the interferences between the
new resonance and the spin-5/2 resonant contributions ($F_{15}$ and
$D_{15}$) were crucial for the neutron target. We estimated
$|\mu_{\gamma nn^*}|\approx0.1\sim0.2$ for the spin-1/2 cases and
$|\mu_{\gamma nn^*}|\approx0.01\sim0.02$ for the spin-3/2 ones to
reproduce the peak at $E_\mathrm{CM}\approx1675$ MeV. However, it was
not easy to determine the spin and parity of the new resonance
unambiguously within the present framework.  

From these observations, the new resonance can be considered as
$N^*(1/2^{\pm},3/2^+)$ along with the estimated strengths of
$\mu_{\gamma NN^*}$ at best in the present work. This is the same
conclusion as in our previous work, in which the spin-3/2 cases were
not taken into account. Among three possibilities of the spin and
parity of $N^*(1675)$, it is interesting to adopt $J^P = 1/2^+$, since
our reaction study implies a photon coupling which is consistent with
the prediction of the chiral quark-soliton model.  The photon coupling
that vanishes for the proton resonance is a general consequence of
SU(3) flavor symmetry when the resonance is identified as a member
of the antidecuplet pentaquark baryons.  Experimentally, we still
need further information in order to establish that the peak structure
in the deuteron target is the real one from the new resonance.  Once
it will be done, we will be able to make another step forward to
exotic hadron spectroscopy.   

%-------------------------------------------------------
\section*{Acknowledgement}
%-------------------------------------------------------
The authors are grateful to fruitful comments from 
J.~K.~Ahn, Y.~Azimov, J.~Kasagi, V. Kuznetsov, T.~Nakano, H.~Shimizu,
and M.~V.~Polyakov.  The present work is supported by the Korea
Research Foundation Grant funded by the Korean Government(MOEHRD)
(KRF-2006-312-C00507).  The work of K.S.C. is partially supported by
the Brain Korea 21 (BK21) project in Center of Excellency for
Developing Physics Researchers of Pusan National University,
Korea. The work of S.i.N. is supported in part by grant for Scientific
Research (Priority Area No. 17070002) from the Ministry of Education,
Culture, Science and Technology, Japan. 
%-------------------------------------------------------

%-------------------------------------------------------
\section*{Appendix}
%-------------------------------------------------------
\subsection{Rarita-Schwinger vector-spinor}
We can write the RS vector-spinors according to their spin states (3/2
and 5/2) as follows: 
\begin{itemize}
\item Spin 3/2
\begin{eqnarray}
u^{\mu}(p_{2},\frac{3}{2})&=&e^{\mu}_{+}(p_{2})u(p_{2},\frac{1}{2}),
\nonumber\\
u^{\mu}(p_{2},\frac{1}{2})&=&\sqrt{\frac{2}{3}}e^{\mu}_{0}(p_{2})u(p_{2},\frac{1}{2}) 
+\sqrt{\frac{1}{3}}e^{\mu}_{+}(p_{2})u(p_{2},-\frac{1}{2}),
\nonumber\\
u^{\mu}(p_{2},-\frac{1}{2})&=&\sqrt{\frac{1}{3}}e^{\mu}_{-}(p_{2})u(p_{2},\frac{1}{2})
+\sqrt{\frac{2}{3}}e^{\mu}_{0}(p_{2})u(p_{2},-\frac{1}{2}),
\nonumber\\
u^{\mu}(p_{2},-\frac{3}{2})&=&e^{\mu}_{-}(p_{2})u(p_{2},-\frac{1}{2}).
\end{eqnarray}
\item Spin 5/2
\begin{eqnarray}
u^{\mu\nu}(p_2,\frac{5}{2})&=&e^{\mu}_{+}e^{\nu}_{+}u(p_2,\frac{1}{2}),
\nonumber\\
u^{\mu\nu}(p_2,\frac{3}{2})&=&\sqrt{\frac{2}{5}}e^{\mu}_{+}
e^{\nu}_{0}u(p_2,\frac{1}{2})+\sqrt{\frac{1}{5}}e^{\mu}_{+}e^{\nu}_{+}
u(p_2,-\frac{1}{2})+\sqrt{\frac{2}{5}}e^{\mu}_{0}e^{\nu}_{+}u(p_2,\frac{1}{2})
\nonumber\\
u^{\mu\nu}(p_2,\frac{1}{2})&=&\sqrt{\frac{1}{10}}e^{\mu}_{+}
e^{\nu}_{-}u(p_2,\frac{1}{2})+\sqrt{\frac{1}{5}}e^{\mu}_{+}e^{\nu}_{0}
u(p_2,-\frac{1}{2})
\nonumber\\
&+&\sqrt{\frac{2}{5}}e^{\mu}_{0}e^{\nu}_{0}u(p_2,\frac{1}{2})
+\sqrt{\frac{1}{5}}e^{\mu}_{0}e^{\nu}_{+}u(p_2,-\frac{1}{2})+\sqrt{
\frac{1}{10}}e^{\mu}_{-}e^{\nu}_{+}u(p_2,\frac{1}{2})
\nonumber\\
u^{\mu\nu}(p_2,-\frac{1}{2})&=&\sqrt{\frac{1}{10}}e^{\mu}_{+}
e^{\nu}_{-}u(p_2,-\frac{1}{2})+\sqrt{\frac{1}{5}}e^{\mu}_{0}e^{\nu}_{-}
u(p_2,\frac{1}{2})
\nonumber\\
&+&\sqrt{\frac{2}{5}}e^{\mu}_{0}e^{\nu}_{0}u(p_2,-\frac{1}{2})
+\sqrt{\frac{1}{5}}e^{\mu}_{-}e^{\nu}_{0}
u(p_2,\frac{1}{2})+\sqrt{\frac{1}{10}}e^{\mu}_{-}e^{\nu}_{+}
u(p_2,-\frac{1}{2})\nonumber\\
u^{\mu\nu}(p_2,-\frac{3}{2})&=&\sqrt{\frac{2}{5}}e^{\mu}_{0}
e^{\nu}_{-}u(p_2,-\frac{1}{2})+\sqrt{\frac{1}{5}}e^{\mu}_{-}
e^{\nu}_{-}u(p_2,-\frac{1}{2})+\sqrt{\frac{2}{5}}e^{\mu}_{-}
e^{\nu}_{0}u(p_2,-\frac{1}{2})
\nonumber\\
u^{\mu\nu}(p_2,-\frac{5}{2})&=&e^{\mu}_{-}e^{\nu}_{-}
u(p_2,-\frac{1}{2}).
\end{eqnarray}
\end{itemize}
Here, we employ the basis four-vectors $e^{\mu}_{\lambda}$ which reads:
\begin{eqnarray}                
e^{\mu}_{\lambda}(p_{2})&=&\left(\frac{\hat{e}_{\lambda}\cdot
{\bm p}_2}{M_{B}},\hat{e}_{\lambda}
+\frac{{\bm p}_{2}(\hat{e}_{\lambda}\cdot{\bm p}_{2})}{M_{B}(
p^{0}_{2}+M_{B})}\right),
\nonumber\\
\hat{e}_{+}&=&-\frac{1}{\sqrt{2}}(1,i,0),\,\hat{e}_{0}=(0,0,1),\,
\hat{e}_{-}
=\frac{1}{\sqrt{2}}(1,-i,0). 
\end{eqnarray}  
%-------------------------------------------------------
 
%-------------------------------------------------------
%FIGURE>>>
\begin{figure}[t]
\includegraphics[width=15cm]{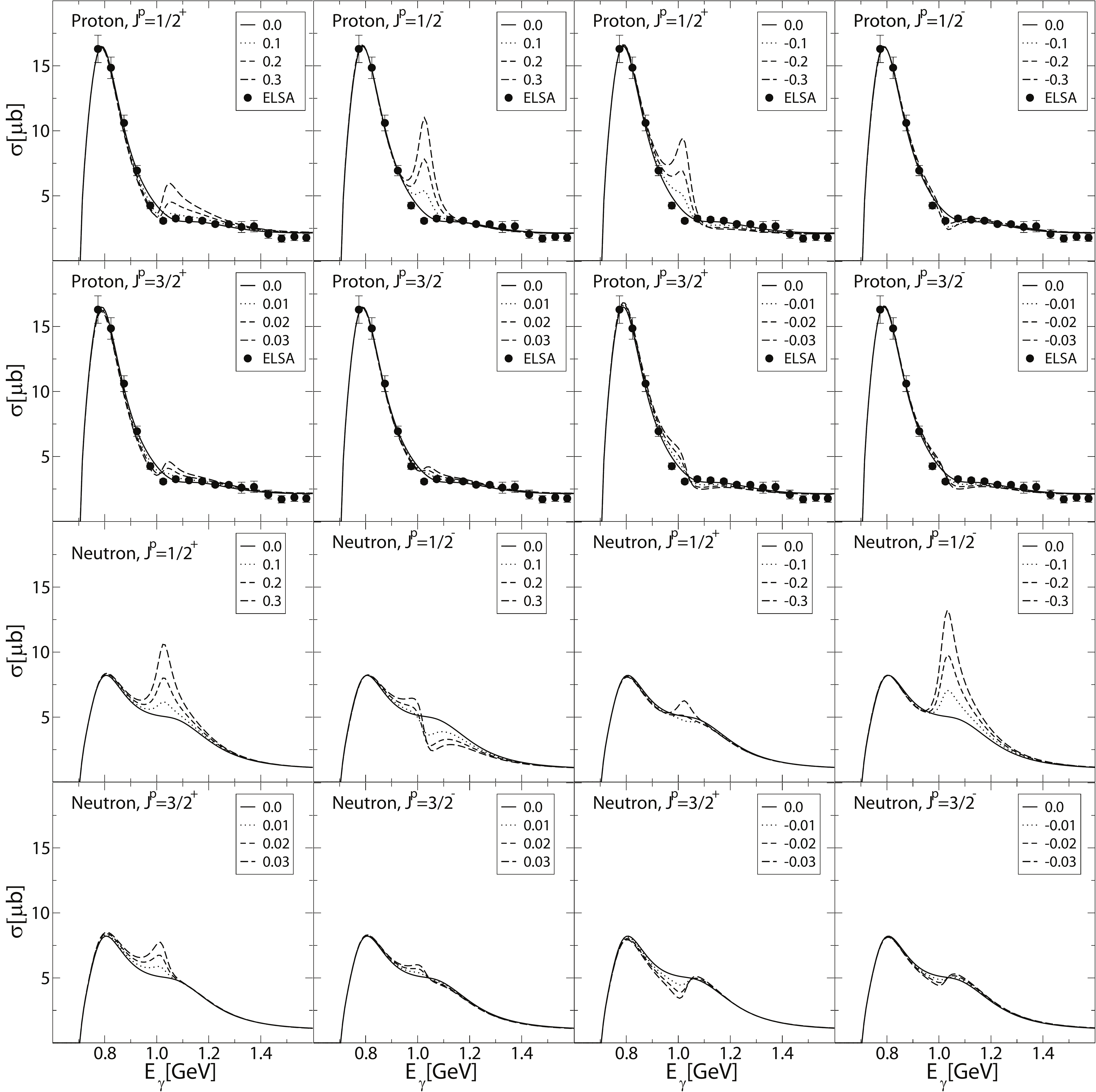}
\label{TC}
\caption{Total cross sections for the proton (first and second rows) and neutron (third and fourth rows) as functions of the photon-laboratory energy in the laboratory frame.}
\end{figure}
%FIGURE<<<
%-------------------------------------------------------
%FIGURE>>>
\begin{figure}[t]
\includegraphics[width=15cm]{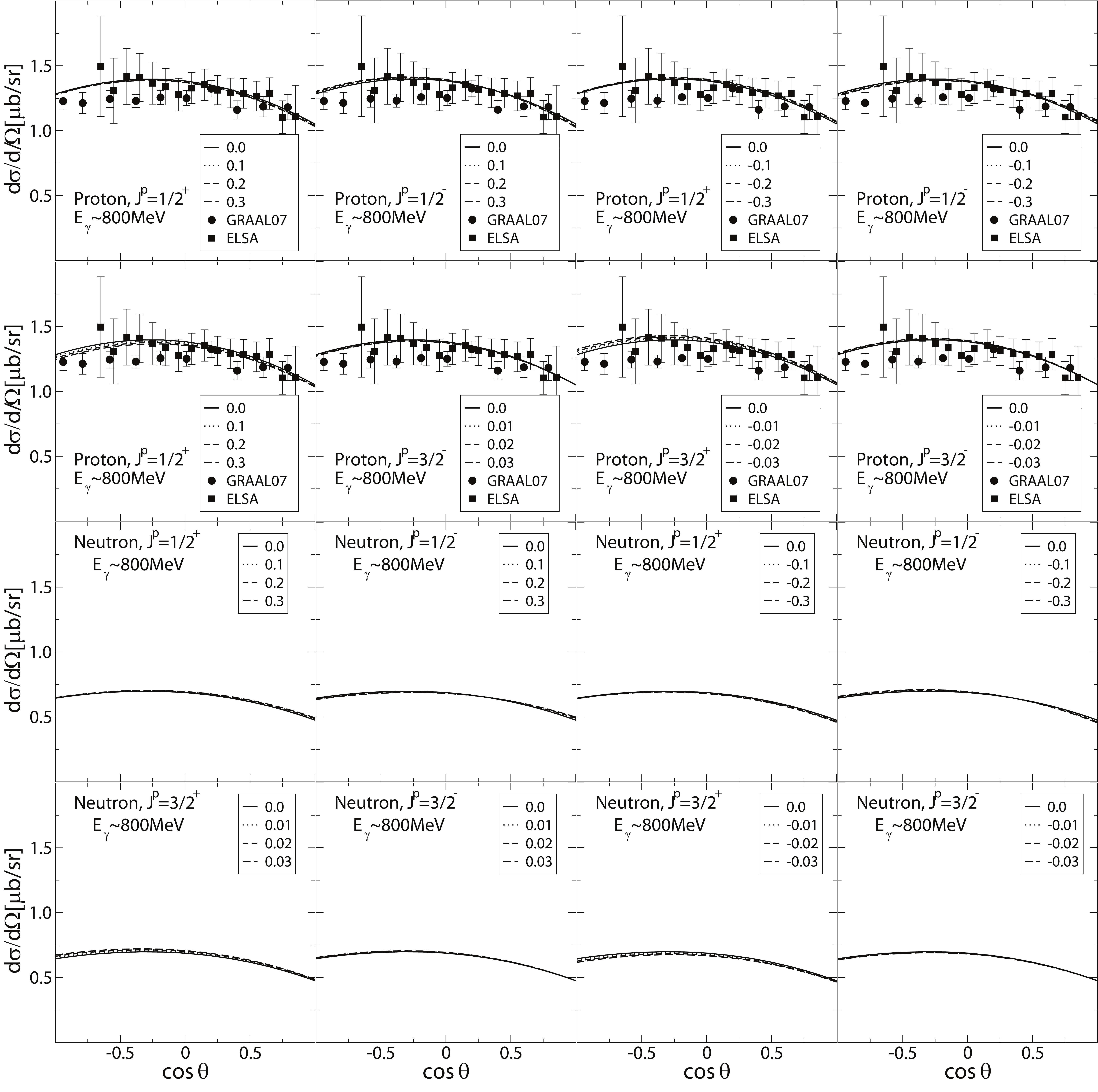}
\label{DCa1}
\caption{Differential cross sections as functions of the scattering angle in the center of mass system $\theta$ at $E_{\gamma}\approx800$ MeV for the proton (first and second rows) and neutron (third and fourth rows).}
\end{figure}
%FIGURE<<<
%-------------------------------------------------------
%FIGURE>>>
\begin{figure}[t]
\includegraphics[width=15cm]{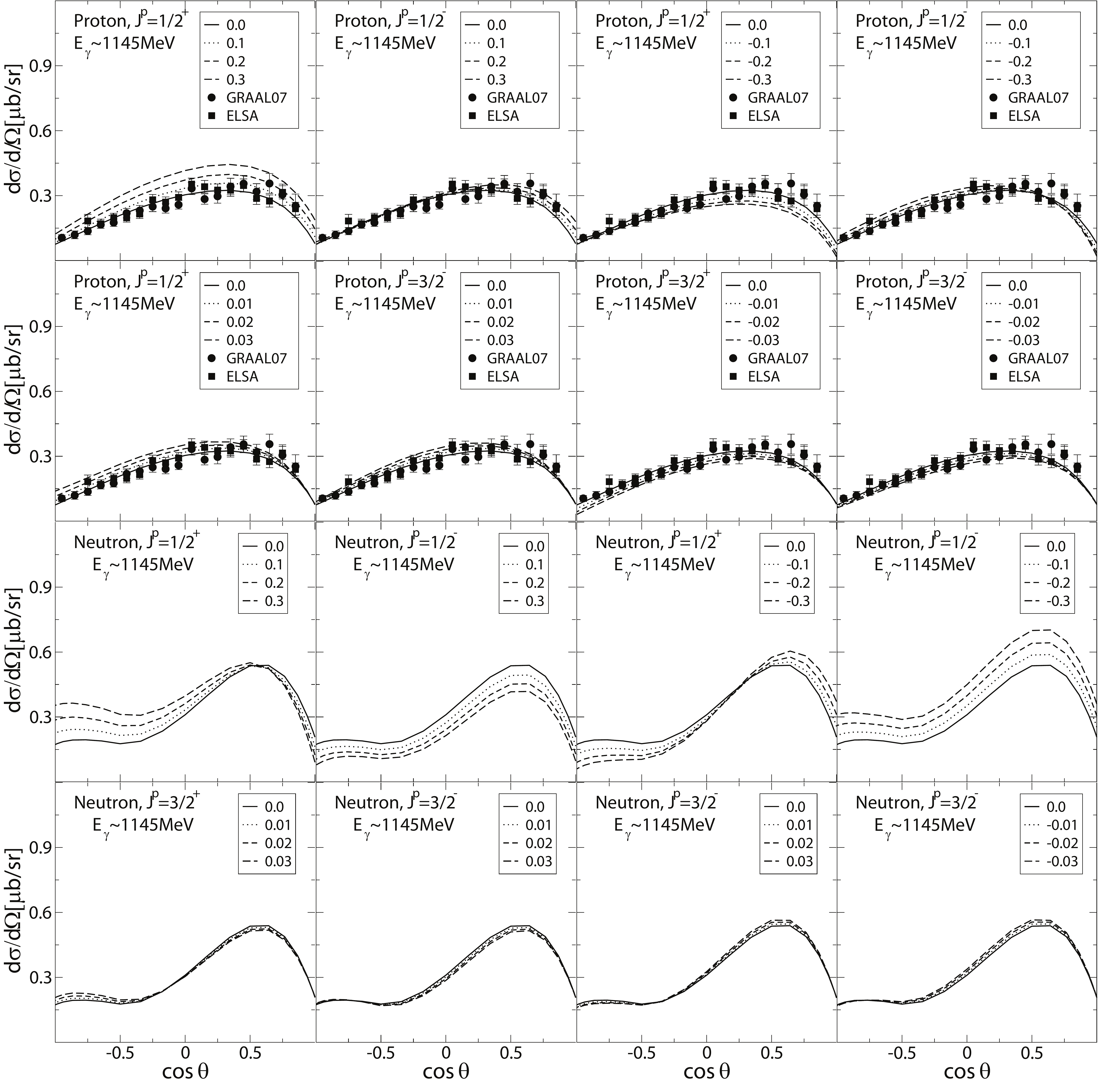}
\label{DCa2}
\caption{Differential cross sections as functions of the scattering angle in the center of mass system $\theta$ at $E_{\gamma}\approx1145$ MeV for the proton (first and second rows) and neutron (third and fourth rows).}
\end{figure}
%FIGURE<<<
%-------------------------------------------------------
%FIGURE>>>
\begin{figure}[t]
\includegraphics[width=15cm]{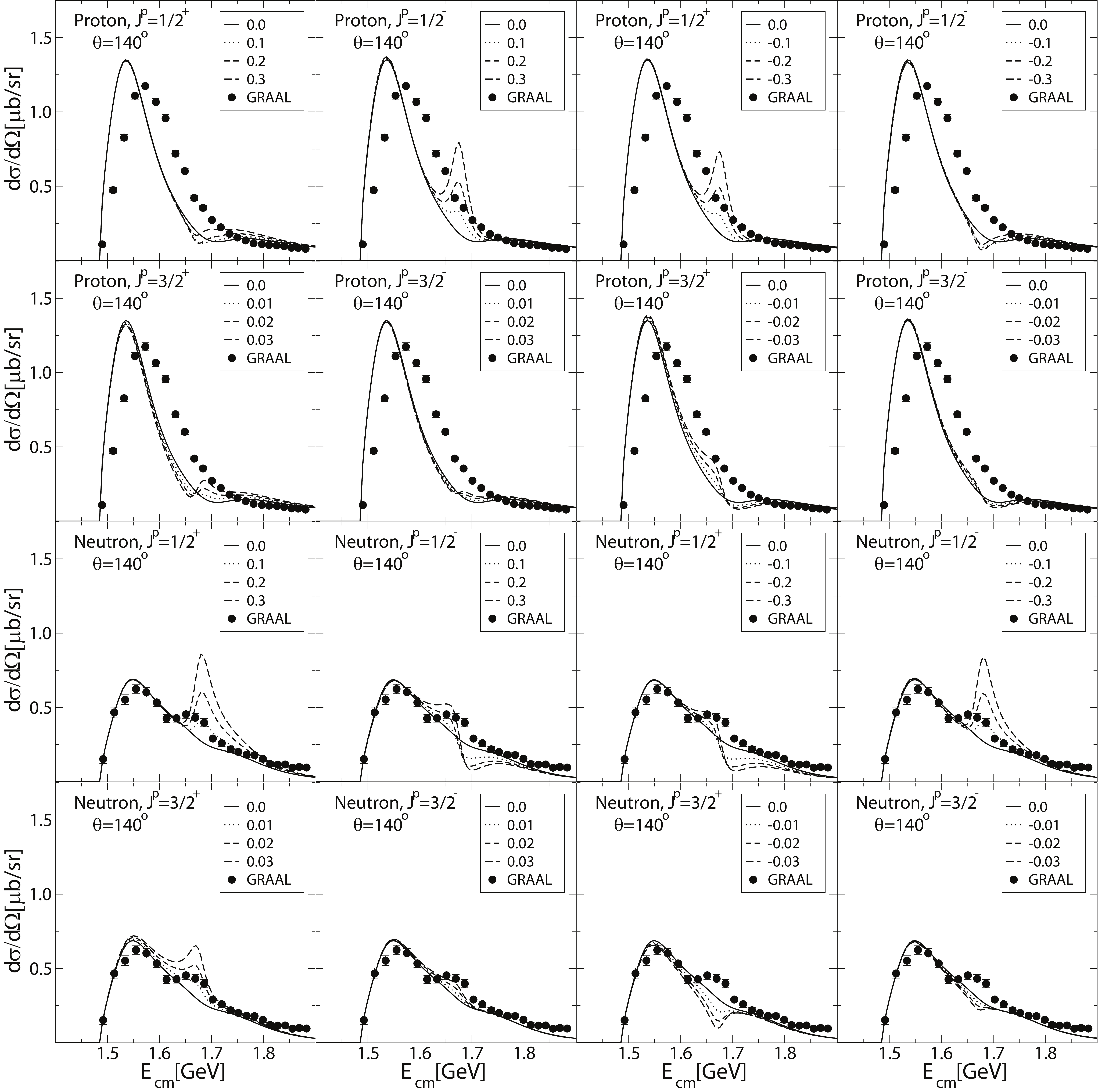}
\label{DCb1}
\caption{Differential cross sections as functions of the photon energy at $\theta\approx65^\circ$ for the proton (first and second rows) and neutron (third and fourth rows).}
\end{figure}
%FIGURE<<<
%-------------------------------------------------------
%FIGURE>>>
\begin{figure}[t]
\includegraphics[width=15cm]{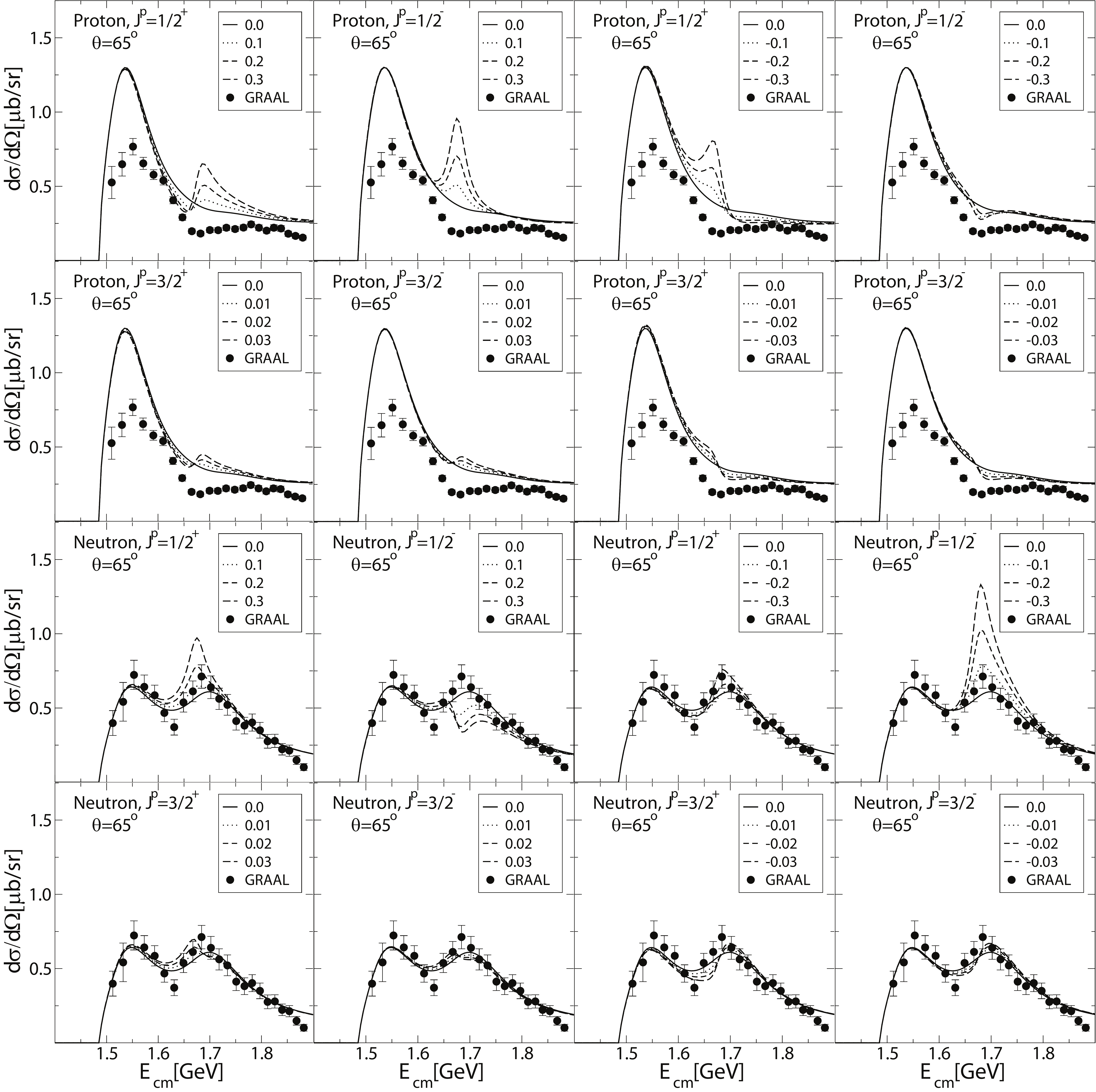}
\label{DCb2}
\caption{Differential cross sections as functions of the photon energy at $\theta\approx140^\circ$ for the proton (first and second rows) and neutron (third and fourth rows).}
\end{figure}
%FIGURE<<<
%-------------------------------------------------------
%FIGURE>>>
\begin{figure}[t]
\includegraphics[width=15cm]{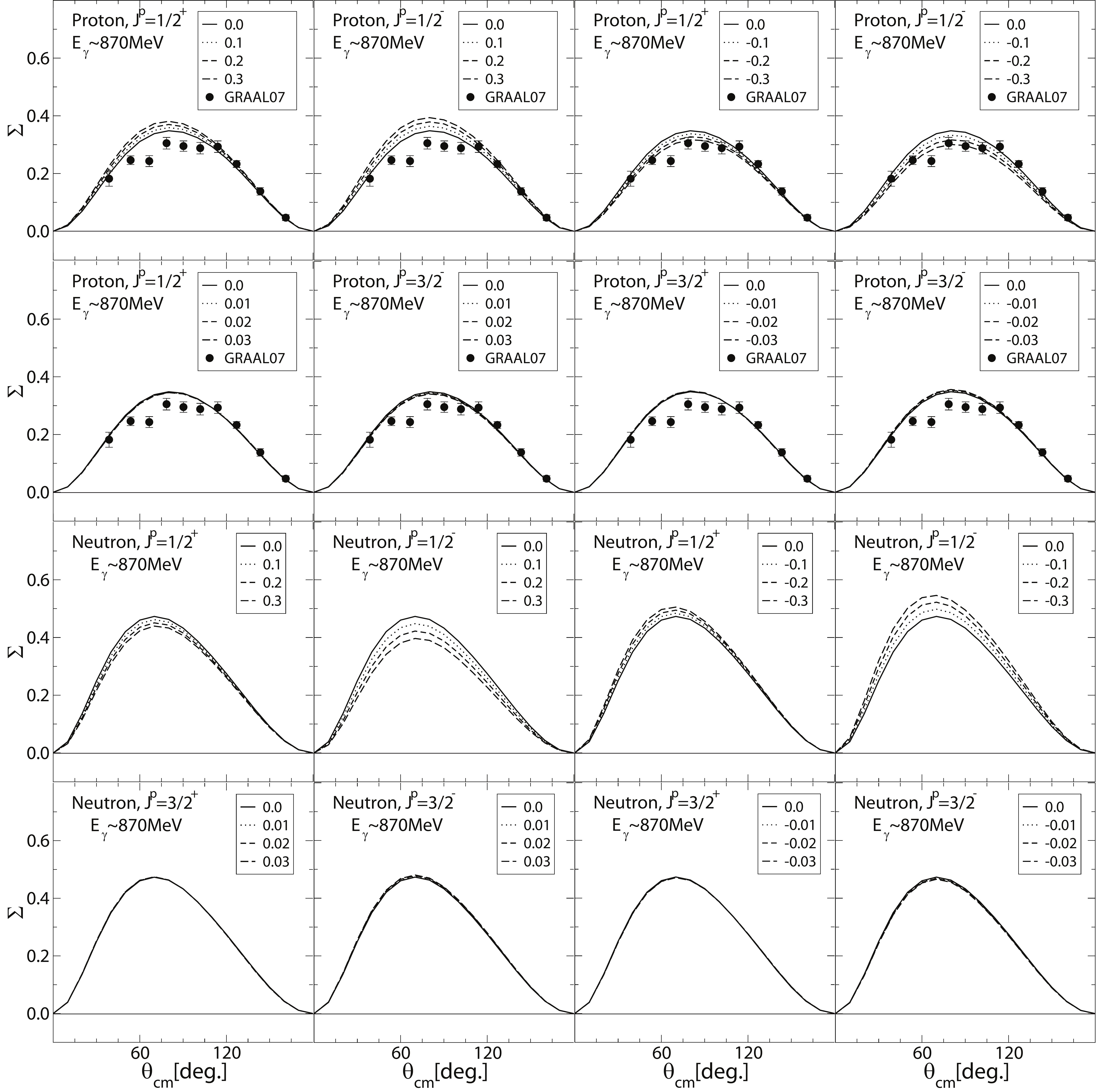}
\label{BA1}
\caption{Beam asymmetries as functions of the scattering angle in the center of mass system $\theta$ at $E_{\gamma}\approx870$ MeV for the proton (first and second rows) and neutron (third and fourth rows).}
\end{figure}
%FIGURE<<<
%-------------------------------------------------------
%FIGURE>>>
\begin{figure}[t]
\includegraphics[width=15cm]{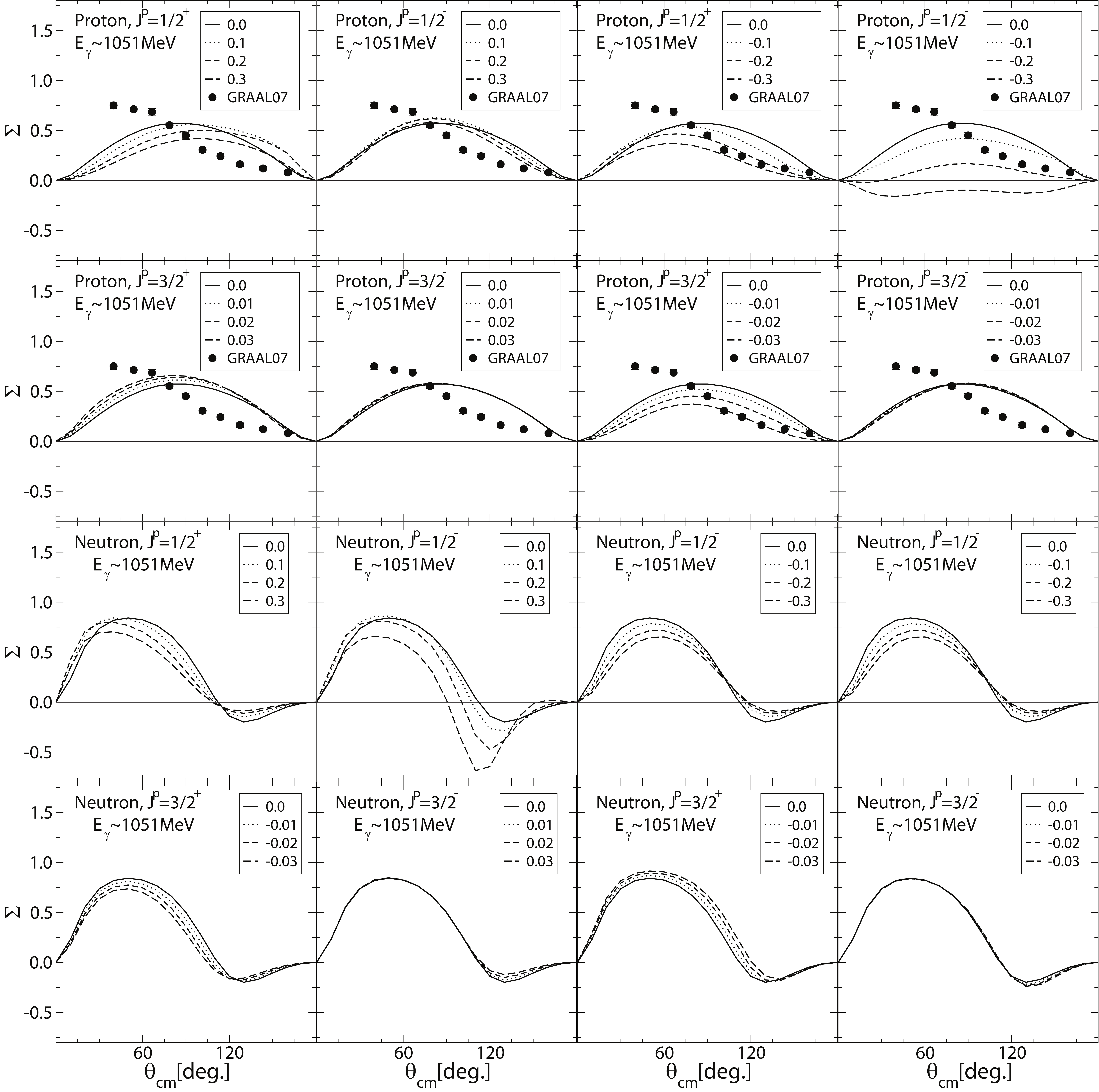}
\label{BA2}
\caption{Beam asymmetries as functions of the scattering angle in the center of mass system $\theta$ at $E_{\gamma}\approx1051$ MeV for the proton (first and second rows) and neutron (third and fourth rows).}
\end{figure}
%FIGURE<<
%-------------------------------------------------------
\end{document}